\documentstyle[11pt,paspconf,epsf]{article}
\begin{document}

\title{On the Stability of the CMB Autocorrelation Function 
on Cosmological Parameters}

\author{A.A.Melkonian}
\affil{Yerevan Physics Institute, Yerevan 375036, Armenia}

\begin{abstract}
 The  properties  of  the Cosmic  Microwave  Background 
Radiation angular autocorrelation  function  in Friedman
 Universe  with  negative 
curvature $(k=-1)$ are studied. The  dependence  of 
the spectral index of autocorrelation  function  on  the  density 
parameter of the Universe is  studied  numerically,  taking  into 
account the effect of geodesic mixing, occuring  in $k=-1$ curvature 
Universes. 
\end{abstract}

\section{Introduction}

The outstanding importance of the knowledge of the power spectrum of the
CMB is connected with the fact that it 
is expected  to contain direct
information on  the spectrum of initial perturbations responsible
for the formation of the present large scale matter distribution of the
Universe (Peebles 1993, Peacock and Dodds 1994). Usually it is assumed 
that the CMB power spectrum which is observed now besides being influenced
 by Sachs-Wolfe effect, basically has to be the same as at the last scattering 
epoch.
If the Universe has negative curvature, due to the
effect of geodesic mixing as shown by  
(Gurzadyan \& Kocharian 1997, Gurzadyan \& Torres 1997),
the present CMB spectrum has to be modified (flattened) after
the time of the last scattering. The information on the spectrum
is obtained via the angular autocorrelation function. The COBE
data indicate the
spectrum $n=1.0 \pm 0.1$ (Smoot et al 1992, Hancock et al 1997, 
Nordberg \& Smoot 1998 ).

Among the important goals 
of the study of anisotropy of CMB is  to check 
 the inflationary models (Efstathiou 1996) . It is well  known,  that 
inflationary scenario has been suggested to explain several 
essential peculiarities  of 
the Universe, including flatness ($\Omega=1$). Among those predictions 
  the   temperature   autocorrelation 
function $C(\theta, \beta)$  will correspond to    Harrison-Zeldovich 
scale-invariant spectrum of initial fluctuations. 

Below we discuss of the interpretation 
of the properties of $C(\theta, \beta)$, where $\theta$ is the 
modulation  angle 
and $\beta$ the observing beam angle in view  of the peculiarities
of motion of CMB photon beams in Friedmann-Robertson-Walker Universe 
with negative curvature.  
 It was proved (Gurzadyan \& Kocharian 1997)
 that in Universe with negative  curvature 
any spectrum of perturbations at last scattering epoch tends to a 
scale-invariant one  with     time.   
Therefore our aim should be to reconstruct the CMB autocorrelation
function at the last scattering epoch from the present form of that
function.

The   CMB    temperature  anisotropy amplitude
 $\Delta T/T$ can be expanded in a series  of 
spherical harmonics (Efstathiou 1996): 

$$
\Delta T/T= \Sigma_{l,m}a_{l,m}Y_{l,m}(\theta, \phi). 
$$

To be valid this 
decomposition functions have to satisfy the following conditions: 

\begin{itemize}
\item The set of $\{Y_{l,m}\}$ functions has to be complete;
\item  $\{Y_{l,m}\}$ functions  have 
to be orthogonal. 
\end{itemize}

However, these conditions certainly valid in flat (k=0) Riemannian
space, but we do not know the basis of functions
 in the case of hyperbolic (k=-1) Universe. So, 
in  this  case  it  is  impossible  to  investigate the  temperature 
anisotropy by common method of harmonic analysis.. 

This difficulty can be avoided if one applies the  
 methods of theory of dynamical systems:  
 the photon  beam  motion  is  reduced   to  the 
problem of the behaviour of time correlation function of geodesic flow 
on 3-manifold with negative constant curvature.  The  problem  is 
how the properties of geodesics flow depend  on  the  geometrical 
and topological properties of the (k=-1) Universe (Gurzadyan \& Kocharian 1994)

The study of the exponential decay  of time  autocorrelation  function 
leads to the decrease of CMB anisotropy after last scattering epoch  depending 
on the value of the density parameter $\Omega$.

This effect was first 
calculated in (Gurzadyan \& Kocharian 1993).

    Note, that the mixing  concept is a result of a  
 "roughness"  of   the system in the   phase   space,   i.e.   the 
possibility of the investigation of its small, but finite regions
(Lichtenberg \& Lieberman 1984).

This  effect  occurs  only  in the Universe  with  negative 
curvature and disappears  in cases k=0 and k=+1.

\section{Autocorrelation function}

The  consideration  of  the  problem  of  exponential  decay   of 
correlators   of   geodesic   flow   have   rather    interesting 
observational   consequences   concerning   the   properties   of 
temperature  autocorrelation  function  $C(\theta, \beta)$.
  Anosov  (Anosov 1962, 1967) and  others had 
obtained  fundamental results concerning   the 
behaviour of geodesic flow on 3-manifold with negative curvature. 
In particular,  it  was  proved  that  a geodesic  flow  on a closed 
(compact,  without  boundary) manifold  with   constant   negative 
curvature is an Anosov  system,    possessing  the  strongest 
chaotic properties  (mixing of all  degrees)  and  have  positive 
Kolmogorov-Sinai entropy (KSE). Note that systems of  this  type 
are structurally   stable, i.e. at 
 the influence of a small perturbation
is ignored and system will remain an Anosov system
This is an important 
point, since FRW Universe is not absolutely homogeneous and 
isotropic, but is perturbed.

Consider the following correlator:

$$
b(\lambda)= \int_{SM} A_1 \circ f^{\lambda}A_2d \mu - 
\int_{SM}A_1d \mu \int_{SM}A_2 d \mu, 
$$

where $f^{\lambda}$ is a geodesic flow on a space with measure $\mu$. 

The correlation function of geodesic flow $f^{\lambda}$ on 
$SM$ with negative curvature is decreasing by exponential low, i.e.

For  $\exists c > 0$ (Pollicott 1992)

$$
|b(\lambda)| \leq c|b(0)|e^{-h \lambda},
$$

where $h$ is the KS-entropy of the geodesic flow $f^{\lambda}$.

The problem is  to describe the free motion of  photon  on  4D 
space. This problem is solved in (Gurzadyan \& Kocharian 1993) 
and for matter-dominated Universe
the following formula was derived:

\begin{equation}
|C_{\lambda}(\theta, \beta)-1| \leq c|C_{0}(\theta, \beta)-1| \frac{1}{(1+z)^2}
\left[ \frac{ \sqrt{1+z \Omega} + \sqrt{1- \Omega}}{1+ \sqrt{a- \Omega}} \right]^4,
\end{equation}

where $z$ is the redshift of the last scattering epoch, $c$ is constant.

\section{Results}

Thus for any   direction in sky  the 
"measured" temperature tends to the constant mean temperature. 
Since this  effect is model independent and appears only in the case  of 
isotropic and homogeneous Universe with negative  curvature.  It provides
 possibility to getting some information on  the 
curvature of the Universe.

\begin{figure}
\plotone{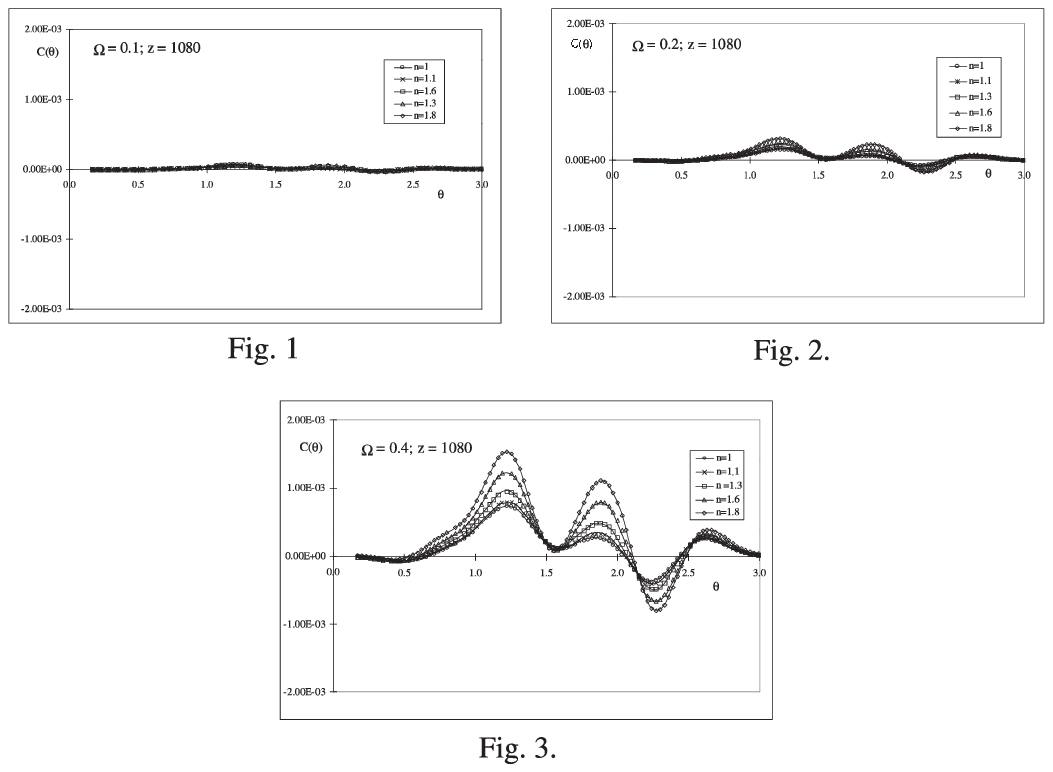}
\end{figure}

Therefore we used (1) to reconstruct the autocorrelation function 
at the last scattering epoch. Calculations were performed for 
several values of the spectral index: $n=1, 1.1, 1.3, 1.6, 1.8$ and
 $z=1000, 1080, 1150$.The results are shown in Figures 1-3. 
The dependence on density parameter $\Omega$ is clearly visible.
\acknowledgments
This work is the fruit of a collaboration with D.Langlois from 
Observatory of Paris-Meudon via French-Armenian Jumelage. 
I am thankful to D.Langlois and 
V.Gurzadyan for valuable discussions and comments.
I am very thankful to Rev.George Coyne for financial support for my visit.

\end{document}